\documentclass[multphys,vecphys]{svmult}

\usepackage{makeidx}         % allows index generation
\usepackage{graphicx}        % standard LaTeX graphics tool
\usepackage{multicol}        % used for the two-column index
\usepackage[bottom]{footmisc}% places footnotes at page bottom
\makeindex             % used for the subject index
\begin{document}

\title*{Tidal Capture by a Black Hole and Flares in Galactic Centres}
\author{Andreja Gomboc\inst{1}, Andrej \v Cade\v z,\inst{1}, Massimo Calvani\inst{2}\and
Uro\v s Kosti\v c\inst{1}}
\institute{Department of Physics, Faculty of Mathematics and Physics, University in Ljubljana, Slovenia
\texttt{andreja.gomboc@fmf.uni-lj.si}
\and INAF, Astronomical Observatory of Padova, Padova, Italy \texttt{calvani@pd.astro.it}}
\maketitle

%\begin{abstract}
%We present the telltale signature of the tidal capture and disruption of an object by a massive black %hole in a galactic centre. As a result of the interaction with the black hole's strong gravitational %field, the object's light curve can flare-up with characteristic time of the order of $\rm{100 s %\times (M_{bh} / 10^6 M_\odot}$). We discuss the two strongest flares that have been observed at the %Galactic centre in 2000 \cite{gom:Baganoff} and in 2003 \cite{gom:Genzel}. Although the first was observed in %X-rays and the second in infrared, they have almost identical light curves and we find it interesting %that it is possible to fit the infrared flare with a rather simple model of the tidally disrupted %comet-like or planetary object. 
%\end{abstract}

The centre of our Galaxy may harbour the nearest (8 kpc) massive black hole. Its proximity allows us to study the environment of massive black holes in detail, including the effects of black hole's gravity on stellar systems in vicinity. 
Recent stellar orbits determinations reveal a central dark mass of $\rm{(3.7\pm 0.2) \times 10^6 \, [R_0/(8 kpc)]^3 \, M_\odot}$, where $\rm{R_0}$ is the distance to Galactic centre \cite{gom:Ghez}. We would like to point out that eccentricities of all these orbits (except one) are close to 1 (see Table 3 in \cite{gom:Ghez}).
In recent years it has been reported that the S0-2 star skimmed the Sgr $\rm{A^*}$ at 17 light hours at the periastron \cite{gom:Schodel}, which corresponds to $\rm{\sim 3000\, r_g}$ (where $\rm{r_g}$ is the gravitational radius of the black hole: $\rm{r_g=GM_{bh}/c^2}$) and the S0-16 came even closer to 45 AU, corresponding to 6.2 light hours \cite{gom:Ghez} or $\rm{\sim 1200 r_g}$ . 

The first rapid X-ray flaring from the direction of Sgr $\rm{A^*}$ was observed in October 2000 with the duration of about 10 ks \cite{gom:Baganoff}. In September 2001, an early phase of a similar X-ray flare was observed in which the luminosity increased by $\approx$ 20 in about 900 s \cite{gom:Goldwurm}. The brightest (observed so far) X-ray flare reaching a factor of 160 of the Sgr $\rm{A^*}$ quiescent value was detected in October 2002 and had a duration of 2.7 ks \cite{gom:Porquet}. In addition, in May and June 2003 four infrared flares from Sgr $\rm{A^*}$ were observed with the duration from $\leq$ 0.9 ks to 5.1 ks and reaching a variability factor of ~5 \cite{gom:Genzel}.

\section{Flares from a tidal disruption of a Solar type star by a $\rm{10^6 M_\odot}$ black hole}
To shed some light on phenomena, which may produce bright flares in galactic centres, we investigate the tidal interaction between a star and massive black hole during a close encounter.  A star approaching the massive black hole will probably follow a highly eccentric orbit.
Once the star plunges deep through the Roche radius:
\begin{equation}
\rm{R_\mathcal{R}=50 \times (\varrho_\odot/\varrho_*)^{1/3} (10^6 M_\odot/M_{bh})^{2/3} r_g,} 
\label{gom-eq:1}
\end{equation}
(where $\rm{\varrho_*}$ and $\rm{\varrho_\odot}$ are star's and Solar density, respectively),
it experiences an enormous work done by tidal forces (reaching as high as $\rm{\sim 0.1 m_*c^2}$) and is disrupted on a timescale of $\rm{\sim 50 r_g/c \sim 250 s \times (M_{bh}/10^6 M_\odot}$). 
As the outer layers of the star are stripped off and the hot core is exposed, the luminosity rises dramatically. The estimates from our numerical simulations show that the rise in luminosity could be as high as 
$\rm{10^{11} - 10^{13} L_\odot}$ (for details see \cite{gom:Gomboc2005} and \cite{gom:Gomboc2001}).
As the stellar debris is scattered and they cool, the luminosity decreases.  Such event would therefore be observed as a bright flare coming from a galactic centre. The exact duration of bright phases depends on the cooling mechanisms and hydrodynamics.

In the case of stars in vicinity of Sgr $\rm{A^*}$, both S0-2 and S0-16 are at their periastrons still safely outside the Roche radius (\ref{gom-eq:1}), which for a Solar type star in the Galactic centre lies at $\rm{R_\mathcal{R} \sim 20 r_g  \sim 7}$ light minutes, and therefore do not get tidally disrupted.

\section{The time scale puzzle of flares in Sagittarius $\rm{A^*}$ and tidal disruption and infall of a comet or asteroid}

In analyzing the flares observed in Sgr $\rm{A^*}$, we were puzzled by the fact that the characteristic rise and switch-off times of all flares are very similar, about 900 sec. Could such a unique time scale suggest common origin for these flares observed at quite different wavelengths?

If the timescale is due to the sources' characteristics, 
they should have almost exactly the same mass. We find this explanation highly unlikely and suggest that the timing is not so much due to sources themselves, but to the space-time of the central Galactic black hole they are moving in.

We explore the idea that observed flares are produced by small objects, e.g. cometary or planetary ones, which are heated by resonant tides on their way down to the black hole. 

In the first phase of such a scenario, the stars moving close to the Galactic centre black hole are gradually being stripped off their comets, asteroids, planets. In the process, the remaining stellar system is loosing its orbital angular momentum, making the stellar system orbit more and more elliptical.

\begin{figure}
\centering
\includegraphics[height=6.8cm]{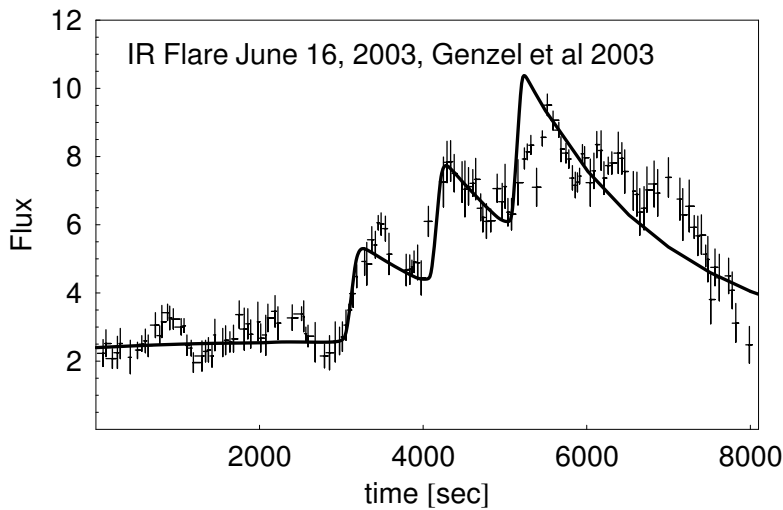}
%\picplace{5cm}{2cm} % Give the correct figure height and width in cm
\caption{Flare observed in Sgr $\rm{A^*}$ and our fit (line) to the observed light curve obtained with a very rudimentary model, assuming that asteroid's luminosity is increasing exponentially with time and the luminosity of its tidal tail is decreasing exponentially with the distance from the asteroid's core.}
\label{gom-fig:1}       % Give a unique label
\end{figure}

In the second phase, a stripped asteroid (with mass $\rm{M}$) is likely to move on a highly eccentric orbit, 
reaching deep into the potential well of the black hole. Each periastron passage produces an increasing tidal wave and reduces the orbital angular momentum and the orbital energy in such a way that the orbit is becoming more and more eccentric (parabolic) with the angular momentum slowly approaching  the angular momentum of tidal capture $\rm{l_{crit}=4 M M_{bh} c}$. 
The last tidal kick, that occurs just before capture, releases up to $\rm{\Delta E \sim 0.1 Mc^2 = 10^{41} erg}$  of tidal energy to the asteroid, which is more than enough to evaporate it and heat it to X-ray temperatures. The result is the formation of a comet-like tidal tail with the length of the circumference of the last circular orbit. The luminosity increases with the characteristic rise time determined by the black hole's gravity: $\rm{\sim 200 s \times (M_{bh} / 10^6 M_\odot}$).We assume that a distant observer is located close to the orbital plane, so that the asteroid's light curve is further modulated by black hole's gravity: as the brightening object is making the last turns before its final demise down the black hole, the Doppler effect, aberration of light, and light bending will produce luminosity peaks with the quasi-period of the last circular orbit - see our fit in Figure \ref{gom-fig:1}. The luminosity in our model decreases, as the object with its tidal tail falls behind the horizon. 

We estimate the capture rate of asteroids as: stellar capture rate $\rm{\times}$ number of asteroids per star, yielding $\rm{\sim (10^{-4} y^{-1}) \times 10^5 = 10 y^{-1}}$.

% Use the \index{} command to code your index words
%\begin{table}
%\centering
%\begin{tabular}{lll}
%\hline\noalign{\smallskip}
%\end{tabular}
%\end{table}

%
%
% BibTeX users please use
% \bibliographystyle{}
% \bibliography{}
%
% Non-BibTeX users please follow the syntax
% the syntax of "referenc.tex" for your own citations
%%%%%%%%%%%%%%%%%%%%%%%% referenc.tex %%%%%%%%%%%%%%%%%%%%%%%%%%%%%%
% sample references
% "physics"
%
% Use this file as a template for your own input.
%
%%%%%%%%%%%%%%%%%%%%%%%% Springer-Verlag %%%%%%%%%%%%%%%%%%%%%%%%%%

%
% BibTeX users please use
% \bibliographystyle{}
% \bibliography{}

\begin{thebibliography}{99.}
% Use the following syntax and markup for your references
%


% Journal
%
\bibitem{gom:Ghez} A.~M.~Ghez, et al.: Astrophys.~J. \textbf{620}, 744 (2005)
%
\bibitem{gom:Schodel} R.~Sch\"{o}del, et al.: Nature. \textbf{419}, 694 (2002)
%
\bibitem{gom:Baganoff} F.~K.~Baganoff, et al.: Nature. \textbf{413}, 45 (2001)
%
\bibitem{gom:Goldwurm} A.~Goldwurm, et al.: Astrophys.~J. \textbf{584}, 751 (2003)
%
\bibitem{gom:Porquet} D.~Porquet, et al.: A\&A. \textbf{407}, L17 (2003)
%
\bibitem{gom:Genzel} R.~Genzel, et al.: Nature. \textbf{425}, 934 (2003)
%
\bibitem{gom:Gomboc2005} A.~Gomboc, A.~\v Cade\v z: Astrophys.~J. \textbf{625}, 278 (2005) 
% Theses
\bibitem{gom:Gomboc2001} A.  Gomboc: Rapid Luminosity Changes Due to Interaction with a Black Hole. PhD
Thesis, University of Ljubljana, Ljubljana (2001)

\end{thebibliography}

%%%%%%%%%%%%%%%%%%%%%%%%%%%%%%%%%%%%%%%%%%%%%%%%%%%%%%%%%%%%%%%%%%%%%%  }
%%%%%%%%%%%%%%%%%%%%%%%%%%%%%%%%%%%%%%%%%%%%%%%%%%%%%%%%%%%%%%%%%%%%%%
\printindex

\end{document}